\documentclass[aps,prd,onecolumn,groupedaddress,nofootinbib]{revtex4}         

\usepackage[english]{babel}

\usepackage[letterpaper,top=2cm,bottom=2cm,left=3cm,right=3cm,marginparwidth=1.75cm]{geometry}

\usepackage{amsmath}
\usepackage{graphicx}
\usepackage[colorlinks=true, allcolors=blue]{hyperref}

\begin{document}


\title{Does Gauss-Bonnet Inflationary Gravitational Waves satisfy the Pulsar Timing Arrays observations?}

\author{Lu Yin}
\email{yinlu@shu.edu.cn}
\affiliation{Department of Physics, Shanghai University, Shanghai, 200444,  China}
\affiliation{Asia Pacific Center for Theoretical Physics, Pohang, 37673, Korea}

\begin{abstract}
The observations from pulsar timing arrays (PTAs), led by the North American Nanohertz Observatory for Gravitational Waves (NANOGrav), have provided opportunities to constrain primordial gravitational waves at low frequencies. In this paper, we analyze the best-fit parameter values for different Gauss-Bonnet Inflationary Gravitational Wave (GB-IGW) models with the  PTArcade program, and we compare the results with the observations of NANOGrav,  European Pulsar Timing Array (EPTA),  Parkes Pulsar Timing Array (PPTA), and International Pulsar Timing Array (IPTA). We find the potential parameter $n$ derived from the GB-IGW model is not necessarily positive. Instead, PTA data suggest the possibility of new parameter ranges. {Meanwhile, the other parameters' results, such as the spectral indices of the scalar and tensor perturbation also have differences with traditional Cosmic Microwave Background analyses from Planck 2018, showing the reason for the different phenomena of Gravitational Wave power spectra expected. Additionally, the GB-IGW models we calculated are closer to the central values of multiple PTA datasets compared to the standard inflation model, making the GB-IGW model more likely to be detected by future space-based gravitational wave observatories.}
\end{abstract}

\maketitle

\section{Introduction}
The theory of cosmic inflation \cite{Guth:1980zm, Albrecht:1982wi, Linde:1981mu} is regarded as a successful early Universe paradigm due to its resolution of major issues inherent in the standard Big Bang cosmological model. This theory also elucidates the scale-invariant spectrum of anisotropies observed in the Cosmic Microwave Background (CMB) and provides the primordial seeds for the large-scale structure of the Universe \cite{WMAP:2008lyn, WMAP:2010qai, Bennett:2010jb, WMAP:2012nax, Planck:2013jfk,  Planck:2015sxf, Planck:2018jri}. Moreover, inflation predicts the emergence of primordial gravitational waves. These gravitational waves (GWs) constitute a stochastic background and propagate largely unimpeded across cosmic distances, thus enabling their detection in the present. 
In contrast, electromagnetic radiation began to propagate freely only after the epoch of recombination \cite{Turner:1993vb, Maggiore:1999vm, Zhao:2006mm, Watanabe:2006qe, Nakayama:2008wy, Kuroyanagi:2008ye, Giovannini:2008tm, Nakayama:2009ce, Kuroyanagi:2011fy, Buchmuller:2013lra, Kuroyanagi:2014nba, Koh:2015brl, Tumurtushaa:2016ars, Caprini:2018mtu, Christensen:2018iqi}. Therefore, the detection of the stochastic gravitational wave background (SGWB) from the inflation will provide a unique and direct way to test the extremely early Universe \cite{Choudhury:2023kam, Choudhury:2023jlt, Choudhury:2023rks, Choudhury:2023hvf, Oikonomou:2024zhs}.

Due to the stretching effect of inflation, the temperature was nearly zero in the early time \cite{Guzzetti:2016mkm, Domenech:2021ztg, Yuan:2021qgz}, and then the Universe entered a reheating phase after inflation ended \cite{Liddle:2000cg, Ashoorioon:2022raz}. During this period, the inflationary field oscillated around its potential minimum and transferred its energy to the plasma of standard model particles \cite{Starobinsky:1979ty, Rubakov:1982df, Fabbri:1983us, Abbott:1984fp}. 
The Gauss-Bonnet term at inflation model we consider offers predictions for different temperatures and equations of state during reheating, resulting in a primordial gravitational wave spectrum that differs from the standard model \cite{Nojiri:2005vv, Satoh:2007gn, Satoh:2008ck, Satoh:2010ep, Guo:2010jr, Nojiri:2010wj, Jiang:2013gza, Koh:2014bka, Koh:2016abf, vandeBruck:2016xvt, Bhattacharjee:2016ohe, Nozari:2017rta, Nojiri:2017ncd, Koh:2018qcy, Chakraborty:2018scm,  Odintsov:2019clh, Nojiri:2023mbo, Nojiri:2023jtf, Biswas:2023eju, Oikonomou:2023qfz, Odintsov:2023aaw, Biswas:2024viz, Oikonomou:2024jqv}. 
The relative parameters of this process —such as the scalar and tensor spectral indices, the tensor-to-scalar ratio, and the $e$-folding number during inflation—can be tested by kinds of observations, such as the $B$-mode polarization in the CMB \cite{Allen:1987bk, Sahni:1990tx, Satoh:2008ck}, or future space-based laser interferometers and pulsar timing experiments.

Several pulsar timing array (PTA) experiments have provided exciting evidence for gravitational wave background signals. For instance, the North American Nanohertz Observatory for Gravitational Waves (NANOGrav \cite{NANOGrav:2023hvm, NANOGrav:2023gor, Agazie:2024stg}) has conducted observations of 68 millisecond pulsars, that exhibit spatial correlations among pulsars consistent with an isotropic gravitational wave background. Moreover, the additional experiments such as the European Pulsar Timing Array (EPTA) \cite{EPTA:2023sfo, EPTA:2023gyr, EPTA:2023xxk, EPTA:2023fyk},  Parkes Pulsar Timing Array (PPTA) \cite{Zic:2023gta}, International Pulsar Timing Array (IPTA) \cite{Perera:2019sca, Antoniadis:2022pcn, IPTA:2023ero}, and Chinese Pulsar Timing Array (CPTA) \cite{Xu:2023wog} have also contributed new dataset. Our paper analyzes the best-fit parameters for the standard model Inflationary Gravitational Wave (IGW), Gauss-Bonnet Inflationary Gravitational Wave (GB-IGW), and a new Gauss-Bonnet Inflationary Gravitational Wave model with exponential tensor perturbation (exp-GB-IGW) obtained by using the PTArcade software \cite{Mitridate:2023oar} and compare with the results from NANOGrav, EPTA, PPTA, and IPTA.

This paper is structured as follows:
The Section \ref{sec:2} provides a review of the basic principles of inflationary models that include a Gauss-Bonnet term. 
In Section \ref{sec:3}, we introduce the gravitational wave background generated by Gauss-Bonnet inflation and discuss the methods for constraining the relevant parameters. 
We present the resulting inflationary gravitational wave power spectra derived from PTAs and discuss the differences in numerical limits of these parameters compared to general CMB constraint results.
Section \ref{sec:4} concludes with a summary of our analysis.

\section{Gauss-Bonnet Inflation}
\label{sec:2}

We start by considering the action of Gauss-Bonnet gravity.  $\xi(\phi)$ is the coupling function shows the Gauss-Bonnet (GB) term coupled to the canonical scalar field $\phi$
\begin{equation}
S=\int d^4 x \sqrt{-g}\left[\frac{1}{2 \kappa^2} R-\frac{1}{2} g^{\mu \nu} \partial_\mu \phi \partial_\nu \phi-V(\phi)-\frac{1}{2} \xi(\phi) R_{\mathrm{GB}}^2\right],
\end{equation}
where the GB term $R_{\mathrm{GB}}$ is involved as
\begin{equation}
R_{\mathrm{GB}}^2=R_{\mu \nu \rho \sigma} R^{\mu \nu \rho \sigma}-4 R_{\mu \nu} R^{\mu \nu}+R^2,
\end{equation}
and the reduced Planck mass can be given as
$\kappa^2=8 \pi G=M_{\mathrm{pl}}^{-2}$.  The Lagrange will not be affected by the GB term if it is back to zero.

Meanwhile, in the Friedmann-Robertson-Walker (FRW) metric, the flat Universe with the scale factor $a$ can be definite by
\begin{equation}
d s^2=-d t^2+a^2\left(d r^2+r^2 d \Omega^2\right),
\end{equation}
where $a(t)$ measures the relative expansion of the Universe. In this way, the background dynamics of this field equations can be modified as 
\begin{equation}
H^2=\frac{\kappa^2}{3}\left(\frac{1}{2} \dot{\phi}^2+V+12 \dot{\xi} H^3\right),
\end{equation}

\begin{equation}
\dot{H}=-\frac{\kappa^2}{2}\left[\dot{\phi}^2-4 \ddot{\xi} H^2-4 \dot{\xi} H\left(2 \dot{H}-H^2\right)\right],
\end{equation}

\begin{equation}
\label{5}
\ddot{\phi}+3 H \dot{\phi}+V_\phi+12 \xi_\phi H^2\left(\dot{H}+H^2\right)=0,
\end{equation}
 where $V_{\phi}$ respect to $\partial V/ \partial \phi$, and $\xi_{\phi} $ equal to $ \partial \xi / {\partial \phi}$. In this paper, the `` $ \dot{} $ " means the derivative to the cosmic time $t$, so we have  $\dot{\xi}$ implies $\xi_{\phi}\dot{\phi}$ and the Hubble parameter $H=\dot{a}/a$. 

{
In nature units, the gravitational wave speed can be calculated as 
\begin{equation}
c_T^2=1-\frac{4\ddot{\xi}-4H\dot{\xi}}{\kappa^{-2}-4H\dot{\xi}}.
\end{equation}
When the different equation $\ddot{\xi}=H\dot{\xi}$, the speed of gravitational waves keep $c_T^2=1$.
For the slow-roll indices, we have the parameters
\begin{equation}
 \quad \delta_1 \equiv 4 \kappa^2 \dot{\xi} H, \quad \delta_2 \equiv \frac{\ddot{\xi}}{\dot{\xi} H}, \quad \delta_3=\frac{\dddot{\xi}}{\dot{\xi} H^2}, \quad
 \epsilon \equiv-\frac{\dot{H}}{H^2}, \quad \eta \equiv \frac{\ddot{H}}{H \dot{H}}, \quad \zeta \equiv \frac{\dddot{H}}{H^2 \dot{H}}.
\end{equation}
We can also replace these parameters by potential and the coupling function as
\begin{equation}
\begin{aligned}
\delta_1 & =-\frac{4 \kappa^2}{3} \xi_\phi V Q \\
\delta_2 & =-\frac{Q}{\kappa^2}\left(\frac{\xi_{\phi \phi}}{\xi_\phi}+\frac{1}{2} \frac{V_\phi}{V}+\frac{Q_\phi}{Q}\right) \\
\delta_3 & =\frac{Q^2}{\kappa^4}\left[\left(\frac{\xi_{\phi \phi \phi}}{\xi_\phi}+\frac{3 \xi_{\phi \phi} V_\phi}{2 \xi_\phi V}+\frac{V_{\phi \phi}}{2 V}\right)+\left(\frac{3 \xi_{\phi \phi}}{\xi_\phi}+\frac{2 V_\phi}{V}\right) \frac{Q_\phi}{Q}+\frac{Q_\phi^2}{Q^2}+\frac{Q_{\phi \phi}}{Q}\right],\\
\epsilon & =\frac{1}{2 \kappa^2} \frac{V_\phi}{V} Q,\\
\eta & =-\frac{Q}{\kappa^2}\left(\frac{V_{\phi \phi}}{V_\phi}+\frac{Q_\phi}{Q}\right) \\
\zeta & =\frac{Q^2}{\kappa^4}\left[\left(\frac{V_{\phi \phi \phi}}{V_\phi}+\frac{V_{\phi \phi}}{2 V}\right)+\left(\frac{3 V_{\phi \phi}}{V_\phi}+\frac{V_\phi}{2 V}\right) \frac{Q_\phi}{Q}+\frac{Q_\phi^2}{Q^2}+\frac{Q_{\phi \phi}}{Q}\right],
\end{aligned}
\end{equation}
where $Q \equiv \frac{V_\phi}{V}+\frac{4}{3} \kappa^4 \xi_\phi V$ in the above calculations.
}

The $e$-folding number quantifies the exponential expansion of the Universe during inflation and helps to explain the large-scale structure of the Universe.
With the above background evolution, the $e$-folding number of GB inflationary expansion can be written as
\begin{equation}
N=\int_{t_*}^{t_{\text {end }}} H d t \simeq \int_{\phi_{\text {end }}}^{\phi_*} \frac{\kappa^2}{Q} d \phi,
\end{equation}
where $Q$ is defined as
 $ Q \equiv \frac{V_\phi}{V}+\frac{4}{3} \kappa^4 \xi_\phi V ,$ and `` $ * $ "  means the moment of a mode $k$ crosses the horizon. So the $e$-folding number can be transformed into a function of potential and GB term.

The scalar and the tensor perturbation during inflation contribute to CMB anisotropies, helping us to understand the Universe's initial conditions and the physics in inflation.
The $ \mathcal{P}_S$ and $ \mathcal{P}_T$  indicate the primordial power spectra of the scalar and tensor perturbation at horizon crossing \cite{Koh:2014bka}.
In such relationship,  we have the spectral indices of the scalar perturbation $n_s-1=d \ln \mathcal{P}_S / d \ln k$,  the spectral indices of the tensor perturbation $n_t=d \ln \mathcal{P}_T / d \ln k$, 
their running spectral indices $\alpha_s=d n_s / d \ln k$, $\alpha_t=d n_t/ d \ln k$, and the tensor-to-scalar ratio $r=\mathcal{P}_T / \mathcal{P}_S$.  
This tensor-to-scalar ratio has been well observed by CMB detectors. We will discuss the constraints of these parameters by present observations, especially the constraints from PTA dataset in the next section.

\section{Primordial Gravitational Waves from Gauss-Bonnet Inflation}
\label{sec:3}

Primordial gravitational waves are thought to have been produced during the early Universe, specifically during cosmic inflation \cite{Battista:2021rlh, Battista:2022hmv, Battista:2022sci, DeFalco:2023djo, Jiang:2023qbm}. These GW background from inflation models carry information about the early time that can help us to understand the fundamental questions in cosmology. The space-based interferometers, such as the LISA \cite{Bartolo:2016ami, Babak:2017tow}, Taiji \cite{Hu:2017mde, Ruan:2018tsw, Wang:2021srv}, and Tianqin \cite{TianQin:2015yph, TianQin:2020hid} are aimed to detect these ripples of spacetime.  The PTA with advanced analysis and future CMB observations are also important to find the primordial gravitational wave signal. We will introduce the calculation of GW background from GB inflation.

\subsection{Gravitational Wave background from Gauss-Bonnet Inflation}
\label{sec:31}

We calculate the power spectra of the SGWB from Gauss-Bonnet Inflation with the potential to be written as
\begin{equation}
\label{v}
V(\phi)=\frac{V_0}{\kappa^4}(\kappa\phi)^n,
\end{equation}
where $V_0$ and $n$ are two dimensionless constant numbers, in the {previous research}, $n>0$ is assumed \cite{Koh:2018qcy}. 
In this paper, we plan to discuss the numerical value of $n$ from the PTA data constraint.
Meanwhile, the coupling function is given by 
\begin{equation}
\label{xi}
\xi(\phi)=\xi_0(\kappa\phi)^{-n}.
\end{equation}

The tensor part of the FRW metric is the point to describe the primordial gravitational wave background. In Fourier space, the tensor perturbation is expanded as
\begin{equation}
h_{i j}(\tau, \mathbf{x})=\sum_\lambda \int \frac{d \mathbf{k}}{(2 \pi)^{3 / 2}} \epsilon_{i j}^\lambda h_{\lambda, \mathbf{k}}(\tau) e^{i \mathbf{k} \mathbf{x}}
\end{equation}
where $h_{ij}$ satisfies the transverse-traceless condition $\partial_i h^{ij}=0$ with $\delta^{ij}h_{ij}=0$, and $\lambda$ is the polarization state of the tensor perturbations. 

The energy density of primordial GW is defined by the tensor part $\rho_{GW}=-T^0_0$. With the spatial average show as bracket, the energy density can be described as
\begin{equation}
\rho_{\mathrm{GW}}=\frac{M_p^2}{4} \int d \ln k\left(\frac{k}{a}\right)^2 \frac{k^3}{\pi^2} \sum_\lambda\left\langle h_{\lambda, k}^{\dagger} h_{\lambda, k}\right\rangle.
\end{equation}

With such energy density, the GW characterization spectra can be definite by the energy spectra
\begin{equation}
\Omega_{\mathrm{GW}}(k)=\frac{1}{\rho_{\text {crit }}} \frac{d \rho_{\mathrm{GW}}}{d \ln k}=\frac{k^2}{12H_0^2}P_T(k),
\end{equation}
where $H_0$ is present value of Hubble parameter, $\rho_{crit}=3H_0^2M_p^2$ is the critical density, and $P_T$ is the power spectra at present time. By considering $k=2\pi a_0 f$ and $a_0=1$, we can also write the GW characterization  spectra as
\begin{equation}
\Omega_{\mathrm{GW}}(f) = \frac{1}{\rho_c} \frac{d \rho_{\mathrm{GW}}(f)}{d \ln f}=\frac{8 \pi^4 f^5}{H_0^2} \frac{\Phi(f)}{\Delta f}
\end{equation}
where $\Phi(f)$ can be considered as timing residual power spectra density with $S(f)=\Phi(f)/\Delta f$. By using $\rho_{\mathrm{GW}}(f)=\sqrt{S/T_{obs}}$,  this $\Omega_{\mathrm{GW}}(f)h^2$ can be compared with the sensitivity of the observations, and $h$ is the dimensionless Hubble constant, that $H_0=100 h $  ${\mathrm{km}} s^{-1}{\mathrm{Mpc}}^{-1}$. 
in the next subsection.

In general, the $P_T$ is the present results of the power spectra, which can be given by the inflationary one $\mathcal{P}_T$ from the transfer function $\mathcal{T}(k)$ as
\begin{equation}
\label{14}
P_T \equiv \frac{k^3}{\pi^2} \sum_\lambda\left\langle h_{\lambda, k}^{\dagger} h_{\lambda, k}\right\rangle=\mathcal{T}^2(k) \mathcal{P}_T(k).
\end{equation}
Considering such tensor perturbation, the inflationary power spectra with the reference pivot scale can be written as
\begin{equation}
\label{15}
\mathcal{P}_T(k)=\mathcal{P}_T\left(k_*\right)\left(\frac{k}{k_*}\right)^{n_t+\frac{\alpha_t}{2} \ln \left(k / k_*\right)},
\end{equation}
where $\mathcal{P}_T\left(k_*\right)$ is the amplitude given by the tensor-to-scalar ratio as $\mathcal{P}_T\left(k_*\right)=r\mathcal{P}_S\left(k_*\right)$. 
The $n_t$ and $\alpha_t$ are the 
spectral indices of the tensor perturbation, which we prepare to get the constraints results by Markov Chain Monte Carlo (MCMC).

Inspired by Eq. \ref{14} and \ref{15}, we consider the running spectral indices will affect the  tensor perturbation with a different exponential form, and modify the  inflationary power spectrum
\begin{equation}
\label{16}
\mathcal{P}_T(k)=\mathcal{P}_T\left(k_*\right)\left(\frac{k}{k_*}\right)^{n_t}e^{ \alpha_t\left(\frac{k_*}{k}-1 \right)^2 }.
\end{equation}
This new function leads to a different evolution history during inflation and reheating. So we name the new GB-IGW model having an exponential tensor perturbation as the exp-GW-IGW model.
This model we consider only based on the phenomenon of PTA observation.
In this work, we will use PTA data to constrain these models in the next subsection.

The transfer function can be given by 
\begin{equation}
\begin{aligned}
\mathcal{T}^2(k)=  \Omega_m^2\left(\frac{3 j_1\left(k \tau_0\right)}{k \tau_0}\right)^2\left(\frac{g_*\left(T_{\mathrm{in}}\right)}{g_{* 0}}\right)\left(\frac{g_{* s 0}}{g_{* s}\left(T_{\mathrm{in}}\right)}\right)^{4 / 3} \mathcal{T}_1^2\left(\frac{k}{k_{\mathrm{eq}}}\right) \mathcal{T}_2^2\left(\frac{k}{k_{\mathrm{th}}}\right),
\end{aligned}
\end{equation}

\begin{equation}
\mathcal{T}_1^2\left(\frac{k}{k_{\mathrm{eq}}}\right)=1+1.65\left(\frac{k}{k_{\mathrm{eq}}}\right)+1.92\left(\frac{k}{k_{\mathrm{eq}}}\right)^2,
\end{equation}

\begin{equation}
\label{eq:gamma}
\mathcal{T}_2^2\left(\frac{k}{k_{\mathrm{th}}}\right)=\left[1+\gamma\left(\frac{k}{k_{\mathrm{th}}}\right)^{\frac{3}{2}}+\sigma\left(\frac{k}{k_{\mathrm{th}}}\right)^2\right]^{-1},
\end{equation}
where $T_{in}$ is the temperature of the Universe reenters horizon, 
$g_*$ is the relativistic degrees of freedom,
$g_{*s}$ is the effective number of light species for the entropy, 
$j_1$ is the first spherical Bessel function,
$k_{eq}$ is the scale of comoving wave numbers at matter and radiation equivalent time, 
$k_{th}$ is the scale of comoving wave numbers at the completion of reheating.
The value of the coefficients $\gamma$ and $\sigma$  in Eq. \ref{eq:gamma} depend on different inflation models. At here,  we select $\gamma \approx -0.23 $ and $\sigma \approx 0.58$.

Therefore, the characteristic power spectrum of GWs from the GB-inflation model can be described as:
\begin{equation}
\label{20}
\begin{aligned}
h_0^2 \Omega_{\mathrm{GW}}= & \frac{3 h_0^2}{32 \pi^2 H_0^2 \tau_0^4 f^2} \Omega_m^2 \mathcal{T}_1^2\left(\frac{f}{f_{\mathrm{eq}}}\right)\mathcal{T}_2^2\left(\frac{f}{f_{\text {th }}}\right) r \mathcal{P}_S\left(\frac{f}{f_*}\right)^{n_t+\frac{\alpha_t}{2} \ln \left(f / f_*\right)},
\end{aligned}
\end{equation}
Meanwhile, with the exponential form of GB inflation tensor perturbation, the characteristic power spectrum of GWs in the exp-GB-IGW model can be shown as:
\begin{equation}
\label{21}
\begin{aligned}
h_0^2 \Omega_{\mathrm{GW}}= & \frac{3 h_0^2}{32 \pi^2 H_0^2 \tau_0^4 f^2} \Omega_m^2 \mathcal{T}_1^2\left(\frac{f}{f_{\mathrm{eq}}}\right)\mathcal{T}_2^2\left(\frac{f}{f_{\text {th }}}\right) r \mathcal{P}_S\left(\frac{f}{f_*}\right)^{n_t}e^{ \alpha_t\left(\frac{f_*}{f}-1 \right)^2 }.
\end{aligned}
\end{equation}
The two models have the same number of free parameters.

\begin{figure}
\centering
\includegraphics[width=0.65\linewidth]{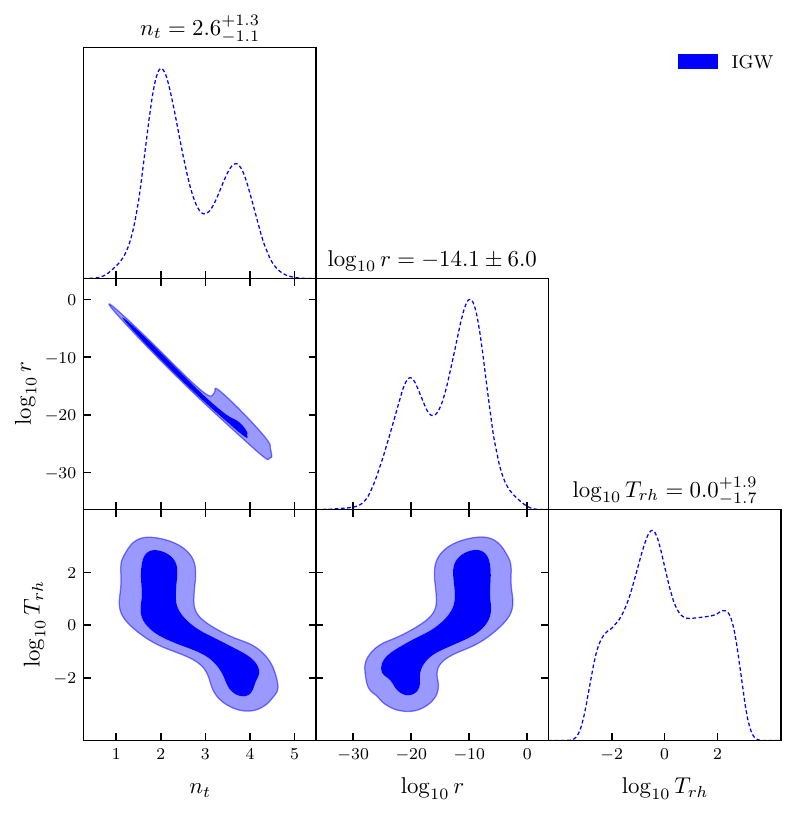}
\caption{\label{fig:IGW}The one and two-dimensional distribution of parameters $n_t$, $\log_{10}r$, $\log_{10} T_{rh}$ in IGW model, where the contour lines represent 68$\%$ and 95$\%$ C.L., respectively. Fitting numerical results in 1D marginalized distributions are shown at the top of the off-diagonal.}
\end{figure}

\begin{figure}
\centering
\includegraphics[width=0.65\linewidth]{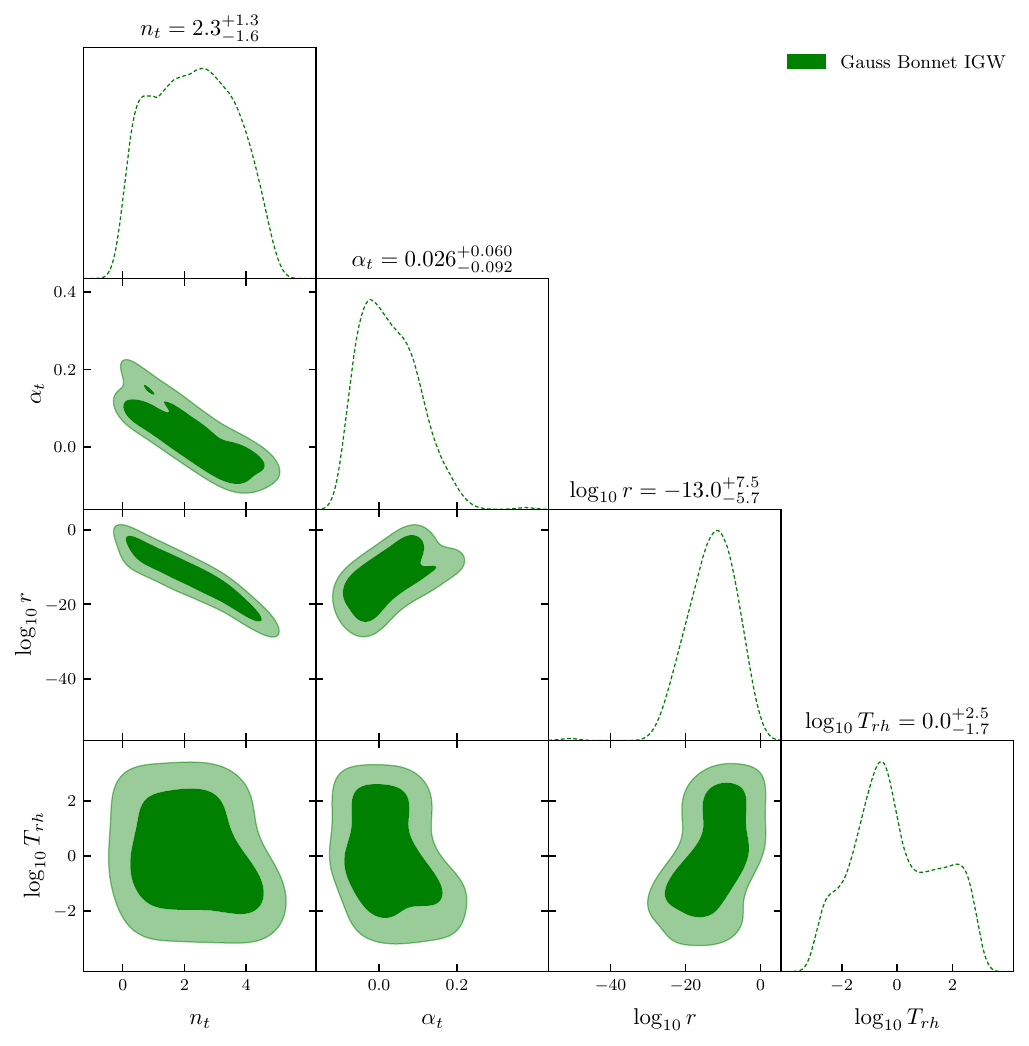}
\caption{\label{fig:IGW-GB}The one and two-dimensional distribution of parameters $n_t$, $\alpha_t$, $\log_{10}r$, $\log_{10} T_{rh}$ in GB-IGW model, where the contour lines represent 68$\%$ and 95$\%$ C.L., respectively. Fitting numerical results in 1D marginalized distributions are shown at the top of the off-diagonal.}
\end{figure}

\begin{figure}
\label{fig:nalpha}
\centering
\includegraphics[width=0.35\linewidth]{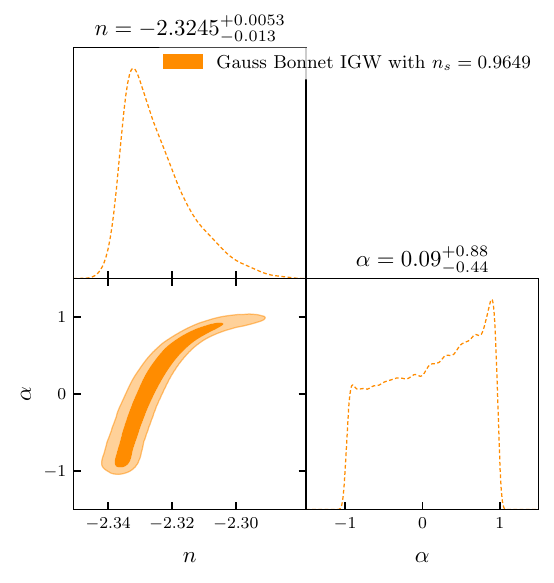}
\includegraphics[width=0.5\linewidth]{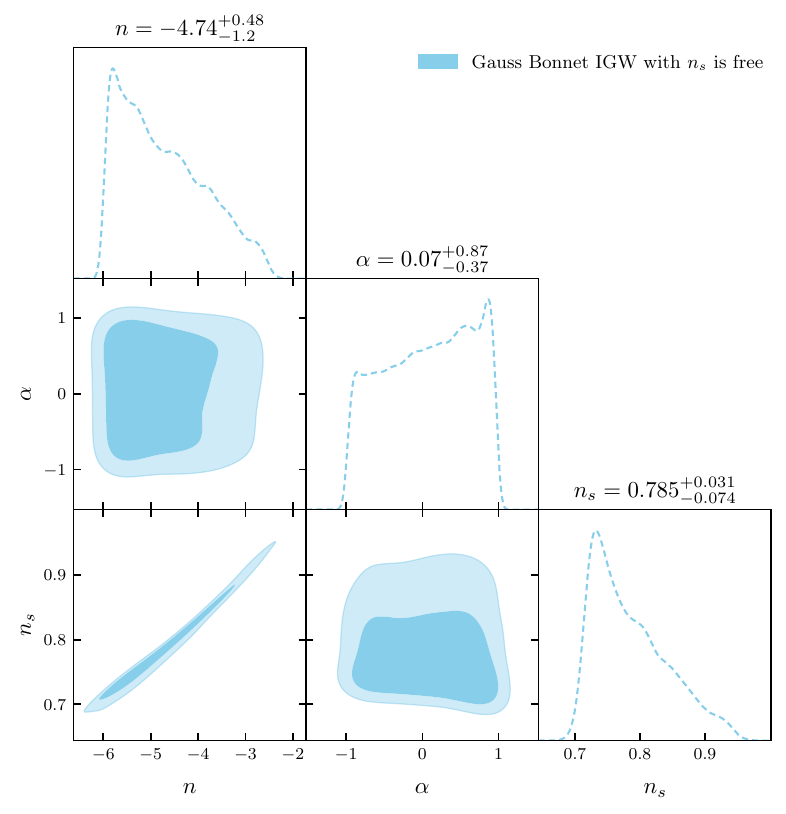}
\caption{\label{fig:IGW-nalpha}The left-hand side figure shows the one and two-dimensional distribution of parameters $n$, $\alpha$ in GB-IGW model with $n_s=0.9649$, where the contour lines represent 68$\%$ and 95$\%$ C.L., respectively. The right-hand side figure shows the one and two-dimensional distribution of parameters $n$, $\alpha$, $n_s$ in the GB-IGW model, where the contour lines represent 68$\%$ and 95$\%$ C.L., respectively. 
Fitting numerical results in 1D marginalized distributions are shown at the top of the off-diagonal. If $\alpha$ is zero, the model back to IGW.}
\end{figure}

\begin{figure}
\centering
\includegraphics[width=0.65\linewidth]{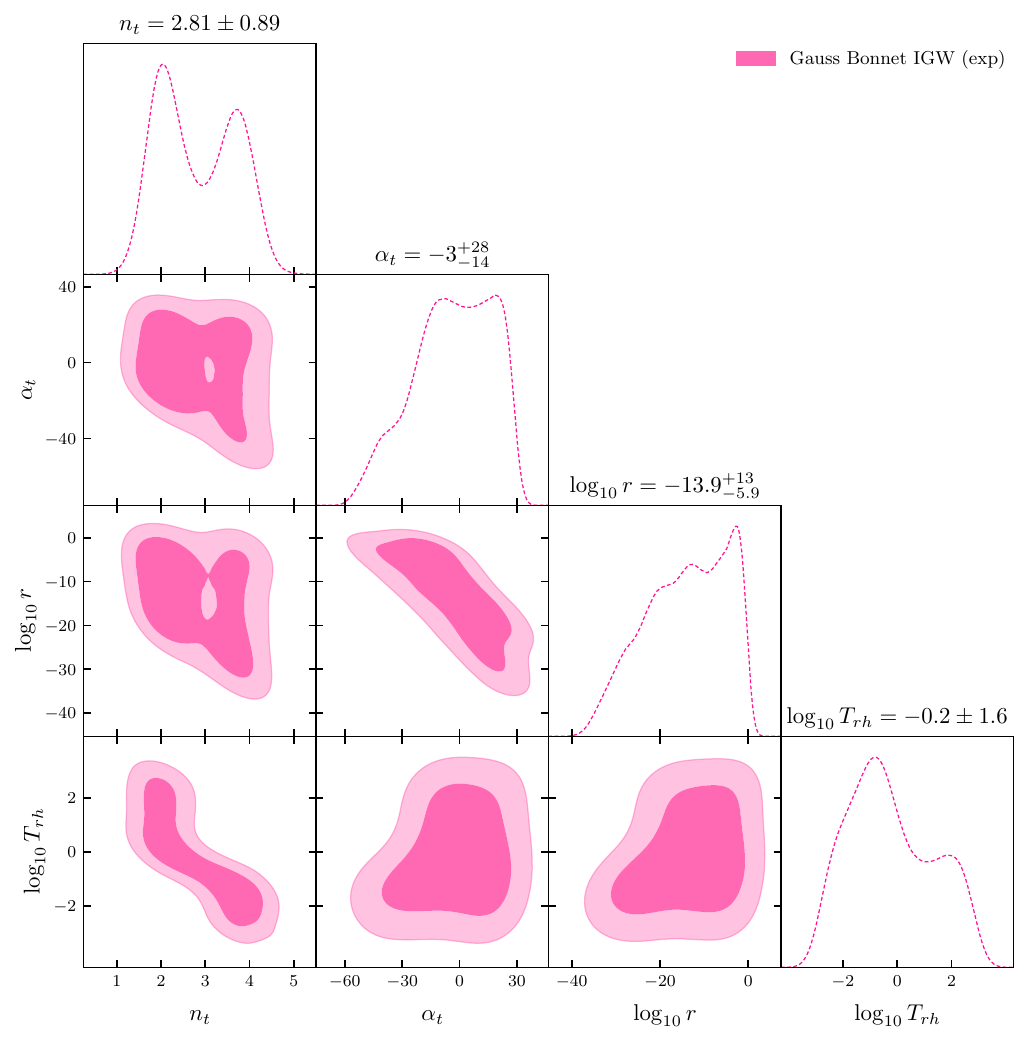}
\caption{\label{fig:IGW-GB-EXP}The one and two-dimensional distribution of parameters $n_t$, $\alpha_t$, $\log_{10}r$, $\log_{10} T_{rh}$ in exp-GB-IGW model, where the contour lines represent 68$\%$ and 95$\%$ C.L., respectively. Fitting numerical results in 1D marginalized distributions are shown at the top of the off-diagonal.}
\end{figure}

\begin{figure}
\centering
\includegraphics[width=0.95\linewidth]{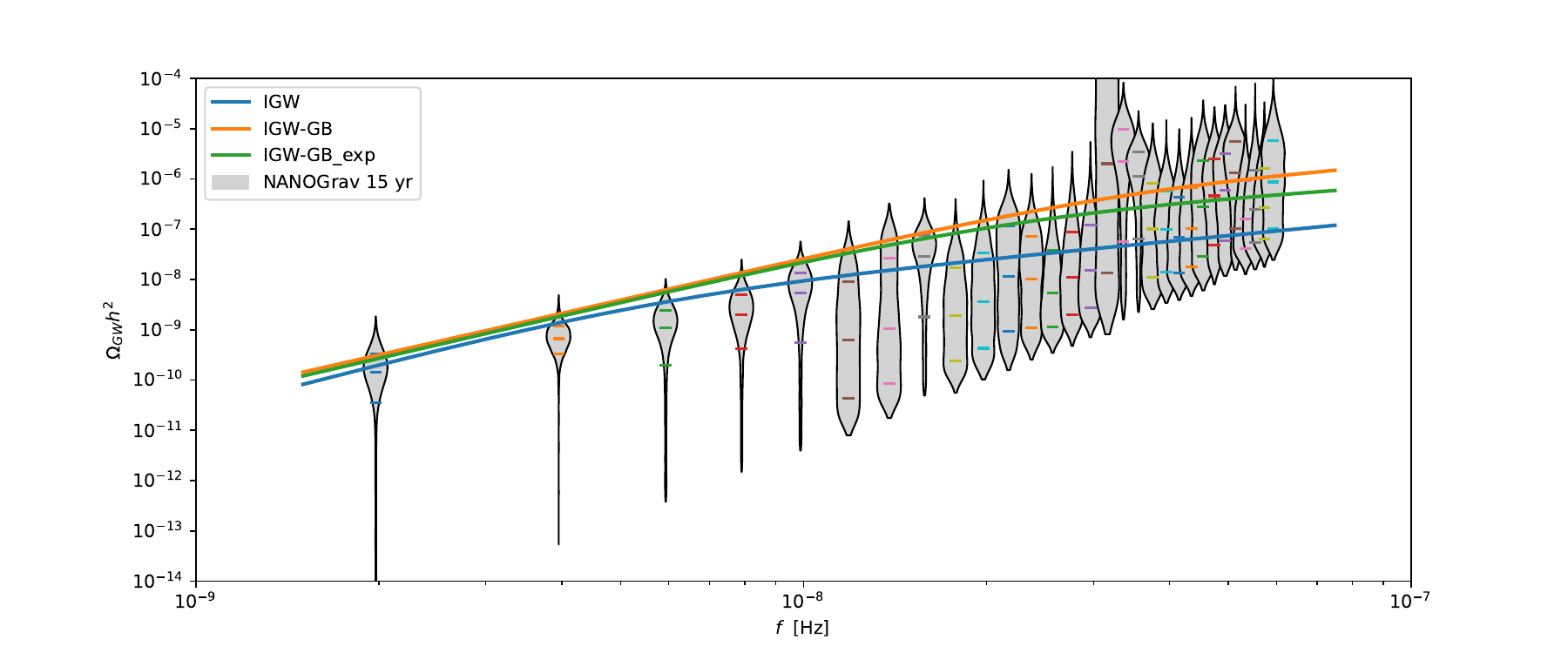}
\caption{\label{fig:NANO}The gray violins plots are the free spectral process for NANOGrav 15 years observation. The blue, orange, and green lines are the GW spectra produced by IGW, GB-IGW, and exp-GB-IGW, respectively.}
\end{figure}

\begin{figure}
\centering
\includegraphics[width=0.95\linewidth]{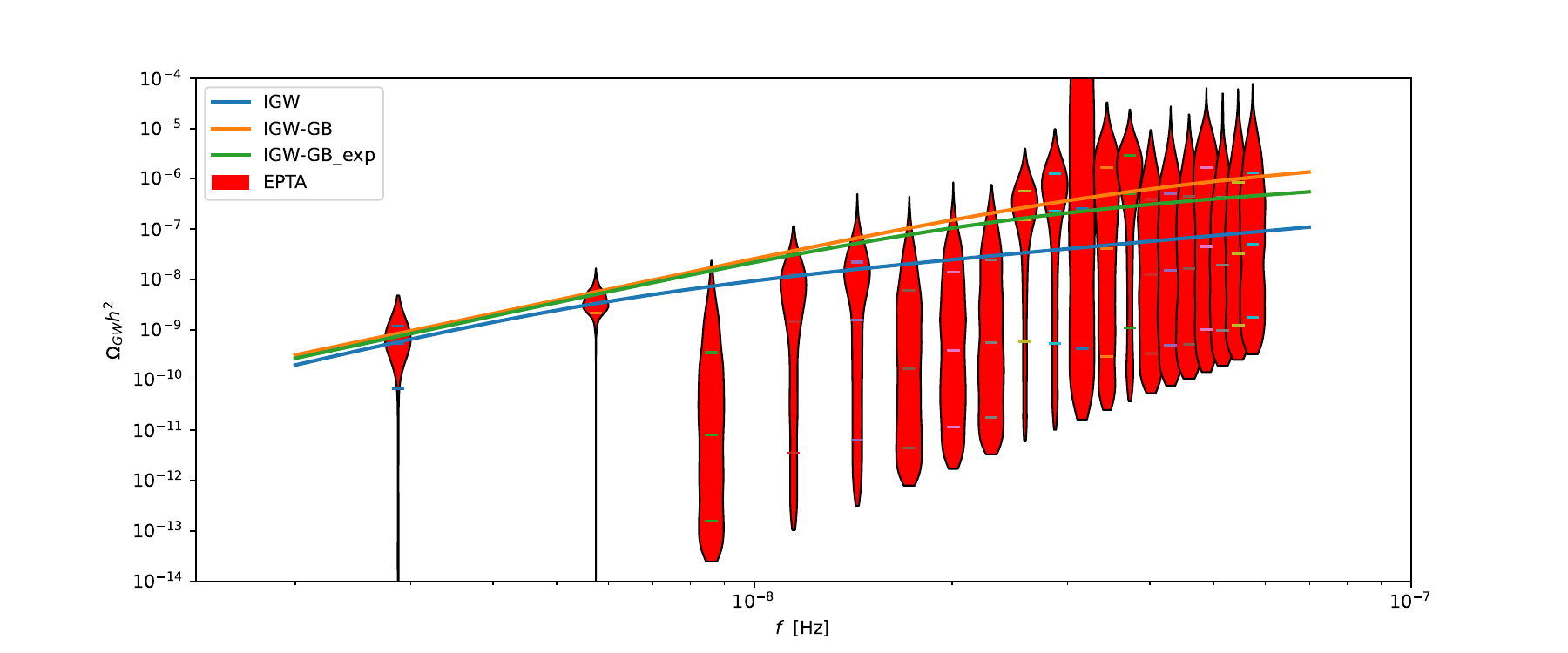}
\caption{\label{fig:EPTA}The red violins plots are the free spectral process for the second data release from the European Pulsar Timing Array observation.  The blue, orange, and green lines are the GW spectra produced by IGW, GB-IGW, and exp-GB-IGW, respectively.}
\end{figure}

\begin{figure}
\centering
\includegraphics[width=0.95\linewidth]{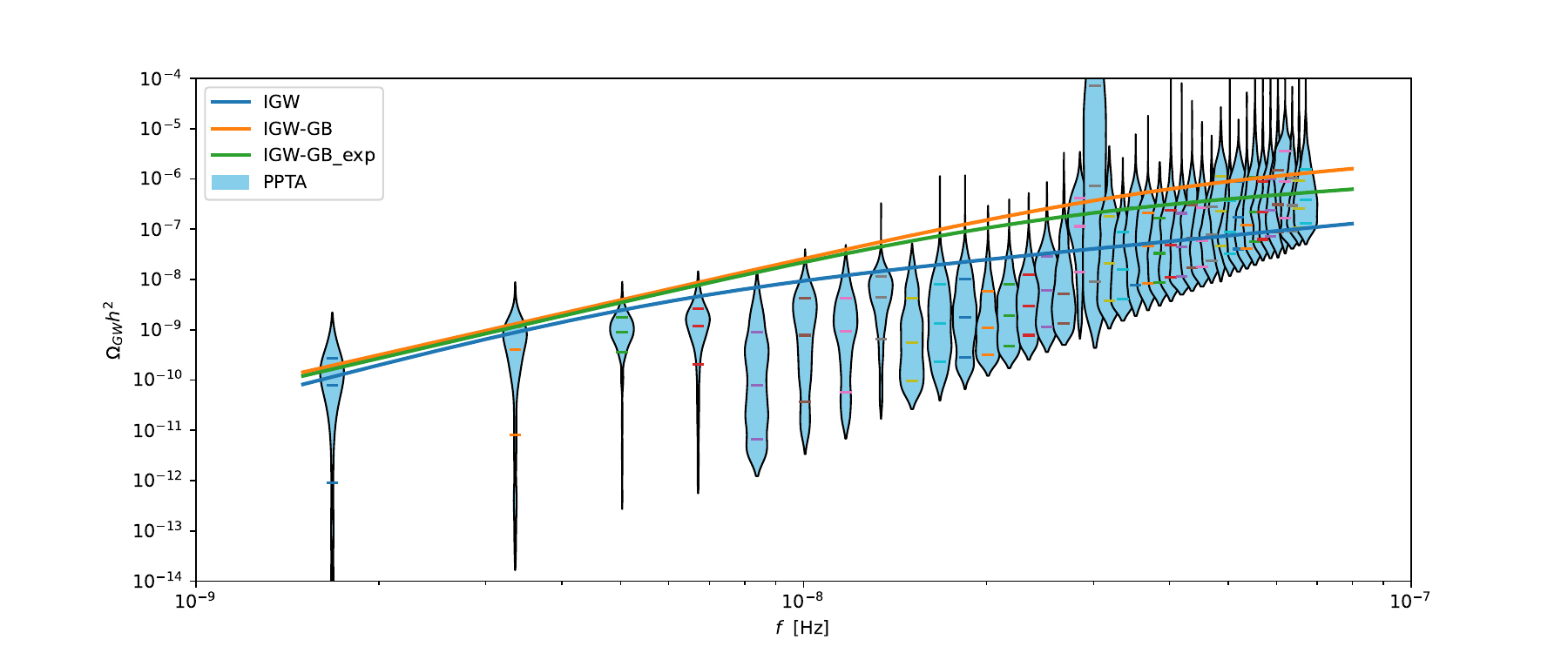}
\caption{\label{fig:PPTA}The sky blue violins plots are the free spectral process for the third data release from the 
Parkes Pulsar Timing Array observation.  The blue, orange, and green lines are the GW spectra produced by IGW, GB-IGW, and exp-GB-IGW, respectively.}
\end{figure}

\begin{figure}
\centering
\includegraphics[width=0.95\linewidth]{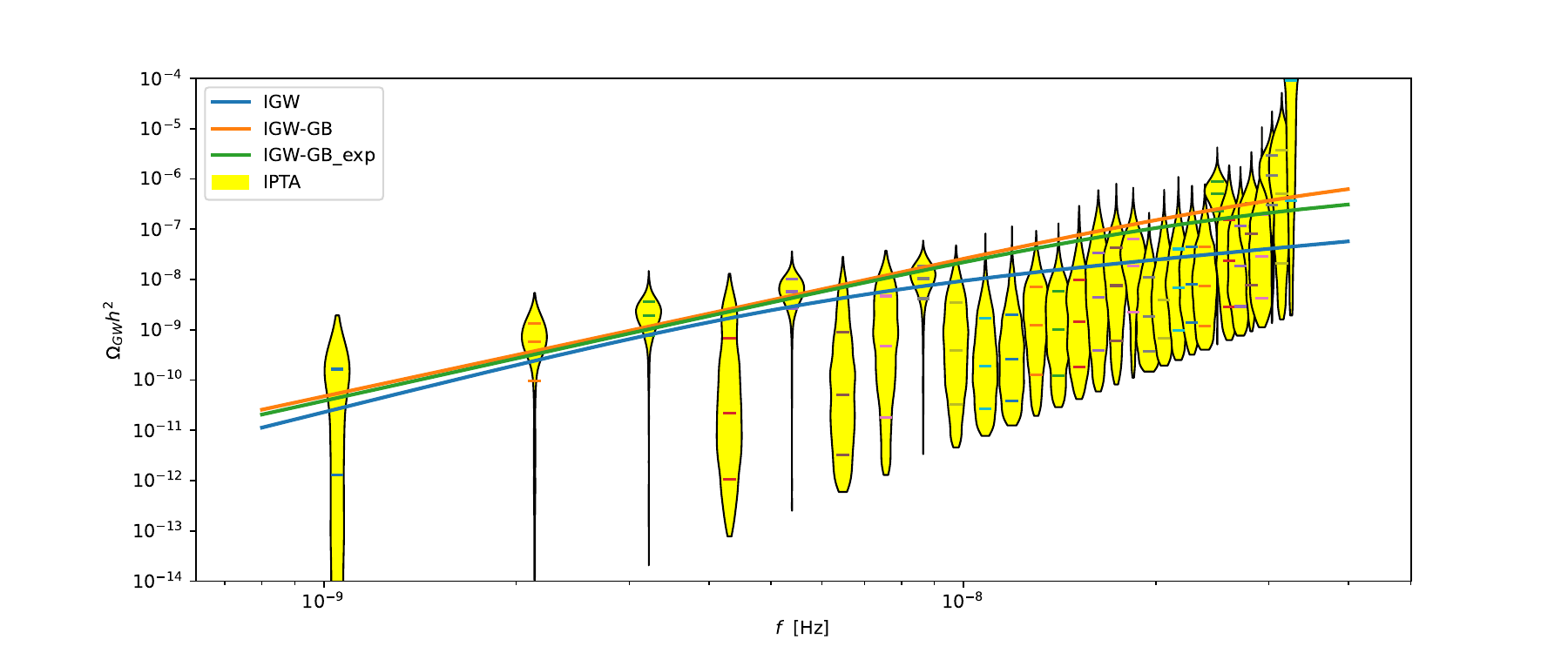}
\caption{\label{fig:IPTA}The yellow violins plots are the free spectral process for the second data release of the International Pulsar Timing Array observation.  The blue, orange, and green lines are the GW spectra produced by IGW, GB-IGW, and exp-GB-IGW, respectively.}
\end{figure}

\subsection{Constrains inflationary GW power spectra from PTA }
\label{sec:32}

In this subsection, we will constrain the parameters of GW power spectra from the GB inflation model, the exp-GB inflation model, and the standard inflation model, respectively. 
The PTArcade \cite{Mitridate:2023oar} software can be used to constrain the free parameters of these models by MCMC. 

We start from the normal inflation theory without the GB term.
The parameter $n_t$ represents the spectral index of the tensor perturbations, $\alpha_t$ denotes the running of the spectral index, $r$ is the tensor-to-scalar ratio, and $T_{rh}$ refers to the reheating temperature. 

Typically, the tensor spectral index $n_t$ is determined by the ``consistency relation".
 However, this conclusion applies only to the standard single-field slow-roll inflation scenario; in non-minimal scenarios, the situation will be changed.
For this fitting analysis, we constrain four parameters to perform a parameterized test of the IGW signal, and not impose the standard consistency relation between $r$ and $n_t$, instead allowing them to vary as independent parameters. Both parameters are permitted to vary independently within broad prior ranges $ \log_{10} r \in [-40, 0] $ and $ n_t \in [0, 6]$. Additionally, we set the range for $ \log_{10} T_{rh} $ as $ [-3, 3] $ with  $T_{rh}$ at the order of GeV.

With MCMC, we obtained the best-fit values for these parameters, as shown in Figure \ref{fig:IGW}. The best-fit value of $\log_{10} r $ is $ -14.1 \pm 6.0 $, { pointing to that PTA data provides a lower numerical result compared to the CMB data from Planck 2018 shows $ r \leq 0.056 $ (95\% confidence level) \cite{Planck:2018jri}. This suggests that detecting the contribution of gravitational waves from NANOGrav 2023 is not only consistent with Planck 2018 in nature but also has higher observational precision in the experiment.
The best-fit value for $n_t$ from NANOGrav 2023 is $ n_t=2.6^{+1.3}_{-1.1} $, which significantly differs from the values traditionally derived from Planck 2018 observations. In the CMB dataset, due to the imposed consistency relation with $n_t = -\frac{r}{8} < 0 $, the resulting $n_t$ is a negative constant. However, PTA data indicates that $n_t$ should be a constant with a positive central value, with a confidence level greater than 2 $\sigma$. This result from NANOGrav 2023 shows the Primordial Gravitational Wave background in the high-frequency should be ``blue-tilted", instead of the ``red-tilted" result given by Planck 2018\cite{Planck:2018jri}.
 $T_{rh} $ decides the temperature of the reheating phase, which in turn affects the thermal history of the Universe. The result of the best-fit value for $ \log_{10} T_{rh} $ is $0^{+1.9}_{-1.7}$. 
 }

Next, we introduce the GB-IGW model, where the parameterized inflationary power spectra for tensor perturbations are influenced by the contributions of $ n_t $ and $\alpha_t$, as shown in Eq. \ref{15}. From this, the formula for the characteristic power spectrum of gravitational waves is derived, as seen in Eq. \ref{20}.
{There are already many works consist Einstein-Gauss-Bonnet theory with the present observations\cite{Khimphun:2016gsn, Koh:2016abf, Koh:2018qcy, Oikonomou:2021kql, Vagnozzi:2023lwo, Oikonomou:2023qfz, Biswas:2023eju, Biswas:2024viz, Oikonomou:2024etl}. Especially, the works by V.K. Oikonomou point out the coupling function of the scalar field and the scalar potential in Gauss-Bonnet invariant cannot be chosen freely\cite{Oikonomou:2021kql, Oikonomou:2023qfz, Oikonomou:2024etl}.
Our fitting results are presented in Figure \ref{fig:IGW-GB}}. The best-fit value for $n_t$ is $2.3^{+1.3}_{-1.6}$, while the optimal value for $\alpha_t$ is $0.026^{+0.060}_{-0.092}$. This non-zero value of $\alpha_t$ indicating its effect is worth considering. Moreover, the positive value of $\alpha_t$ suggests a proportional relationship between $ k $ and $ k_* $. Additionally, we obtained a larger value for $ \log_{10} r = -13^{+7.5}_{-5.7} $, which is larger than the value in the IGW model, implying GB-IGW is easier to be observed by further detectors. 

In previous works about the GB-inflation model, the numerical result of characteristic GW spectra is a line reduce with the increase of frequency \cite{Koh:2018qcy}. This is because the parameter $ n $ in the potential (Eq.\ref{v}) is set to be an integer greater than zero by default. By using the CMB observed result for the spectral index $n_s $ is 0.9649 and a positive value of $n$, the result of  $n_t$ follows the function of given by
 \begin{equation}
n_t=\frac{n(n_s-1)}{n+2}
\end{equation}
 also get a number less than zero, which causes the power spectrum curve to decrease with increasing frequency.

Since $n_t $, $ r $, and $ \alpha_t $ can also be substituted with $ n $, $ \alpha $, and $ n_s $ as, 
\begin{equation}
r=-\frac{8n(n-\alpha)(n_s-1)}{n+2},
\end{equation}
\begin{equation}
\alpha_t=\frac{2n(n_s-1)^2}{(n+2)^2},
\end{equation}
where $\alpha\equiv 4V_0 \xi_0/3 $ is a parameter related to the initial condition of the potential in Eq. \ref{v} and \ref{xi}, and the GB-IGW will back to IGW if $\alpha$ equal to zero.
we once again use the PTA data to fit $n$, $ \alpha $, and $n_s$, treating them as independent parameters.

As shown in Figure \ref{fig:IGW-nalpha}, on the left,  $n_s$ is fixed at 0.9649, while only $n$ and $ \alpha$ are fitted. 
We can see the best-fit value for $n$ is $-2.3245^{+0.0053}_{-0.013}$, which is not the positive constant assumed in previous experimental hypotheses. The result of $ \alpha$ is $0.09^{+0.88}_{-0.44}$, which is also greater than zero, showing the PTA dataset prefers the GB-IGW model than the IGW model. On the right-hand side, $ n $, $ \alpha $, and $n_s$ are fitted simultaneously as free parameters.
The fitting results also show that the best-fit value for $ n$ is $-4.74^{+0.48}_{-1.2}$, not the positive constant assumed in previous experimental hypotheses. $\alpha$ is $0.07^{+0.87}_{-0.37}$, which central value also didn't back to zero, confirming the PTA dataset prefers the GB-IGW model over than IGW model. Additionally, $n_s = 0.785^{+0.031}_{-0.074}$, which is slightly less than the result from CMB observations.

In this paper, we specifically propose a new form of parameterized inflationary power spectra for tensor perturbations exp-GB-IGW model. In this form, $\alpha_t$, $ k $, and $ k_*$ appear in an exponential expression, as shown in Eq. \ref{16}. Under this condition, the formula for the characteristic power spectrum of gravitational waves can also be written as Eq. \ref{21}. The fitting results for this new form are presented in Figure \ref{fig:IGW-GB-EXP}. The best-fit value for $n_t$ is $2.81\pm0.89$, the highest among the three models. The best-fit value for $\alpha_t$ is $-3^{+28}_{-14}$; if it were zero, the model would revert to the standard IGW model. However, based on the results for $r$ and particularly $ T_{rh} $, there is still a significant difference between the best-fit values and those from the IGW model.  The best-fit value of $\log_{10}T_{rh}$ is $-0.2\pm1.6$, indicating the lower temperature of reheating time than the standard model and altering the rate of early structure formation.

After discussing the MCMC results, we now compare the best-fit values of these three models with the data obtained from various PTA observations. 
First, we examine the results from NANOGrav 15,  at high frequencies, the power spectrum of the IGW model is closer to the lower bound of the data, while the power spectra of the GB-IGW and exp-GB-IGW models are relatively higher. Among these, the exp-GB-IGW model is closer to the central point of the data and exhibits a smaller $\chi^2$ value compared to the GB-IGW model. Due to their higher sensitivity, the GB-IGW and exp-GB-IGW models are also more likely to be detected by future space-based observatories than the IGW model.

We also examined the differences between the GW characteristic power spectra provided by the three models and various PTA observations, including EPTA, PPTA, and IPTA. 
The red shaded region in Figure \ref{fig:EPTA} represents the data from EPTA, which is collected by the second data release from the European Pulsar Timing Array \cite{EPTA:2023sfo, EPTA:2023gyr, EPTA:2023xxk, EPTA:2023fyk}, 
showing a broader power spectrum range at high frequencies compared to NANOGrav. 
The blue-shaded region in Figure \ref{fig:PPTA} represents PPTA data, which is obtained by 
the third data release from the 
Parkes Pulsar Timing Array project \cite{Zic:2023gta}. The PPTA dataset exhibits a wider frequency range and more data points than both EPTA and NANOGrav. Notably, at a frequency of $10^{-8}$, the IGW model is closer to the central value of the data, while the GB-IGW and exp-GB-IGW models also fall within the detection range of PPTA. 
At higher frequencies ($10^{-7}$) and lower frequencies ($10^{-9}$), the exp-GB-IGW model is closer to the observed central values compared to the other two models. Finally, IPTA represents the combined results from the second data release of the International Pulsar Timing Array \cite{Perera:2019sca, Antoniadis:2022pcn, IPTA:2023ero}, featuring the lowest frequency range. All three models fall within the range of IPTA data, with the exp-GB-IGW model being closer to the central value at high frequencies. Overall, while the IGW model has a smaller $\chi^2$ value, the GB-IGW and exp-GB-IGW models are closer to the observed central values at both high and low frequencies.

\section{Summary}
\label{sec:4}

In this paper, we analyzed the process of gravitational wave generation from Gauss-Bonnet inflation, calculating the tensor perturbations and the resulting gravitational wave characteristic power spectra. We also explored a new exponential form of the Gauss-Bonnet tensor perturbation named the exp-GB-IGW model and computed its associated gravitational wave characteristic power spectrum. 

By using the PTArcade software and employing MCMC methods, we constrained the model parameters for three models.
We find the present PTA data prefer the value of the potential index $n$ in the Gauss-Bonnet term is negative, which is different from the assumption in previous works. {{ Meanwhile, with considering non-minimal scenarios, the best-fit result shows value of the spectral indices of the tensor perturbation $n_t$ is positive, which is different from the result in CMB observed by Planck 2018. This result is because of the ``blue tilt" observed in PTA, and Planck 2018 gives a ``red tilt" primordial gravitational wave spectra. 
Our fitting result also shows the effect of $\alpha_t$ and $\alpha$ would be non-zero, which means the PTA dataset prefers the GB-IGW model over the standard IGW model. This is one of the important phenomena of the GB-IGW. Moreover, the new model exp-GB-IGW's result of $T_{rh}$ will lead to a lower temperature at the end of reheating.}}

After MCMC, we compared the best-fit results with the power spectra from NANOGrav, EPTA, PPTA, and IPTA.
Our findings reveal that the standard IGW model exhibits a smaller \(\chi^2\), indicating a generally better fit. However, both the GB-IGW and exp-GB-IGW models provide a closer match to the observed central values at both high and low frequencies.  {{We look forward to the next generation of CMB experiments, such as LiteBIRD\cite{LiteBIRD:2022cnt}, which will provide more precise measurements. These future observations will help us determine the exact values of key parameters and further refine the phenomenological constraints on the Gauss-Bonnet model.}}

\section*{Acknowledgements}
We thank William G. Lamb for the help in using NANOGrav data, especially allowed us to use his code have not been published yet. We thank Bum-Hoon Lee, Puxun Wu, and Gansukh Tumurtushaa for the useful discussion. 
L. Yin was supported by an appointment to the YST Program at the APCTP and Natural Science Foundation of Shanghai
24ZR1424600.


\end{document}